# Comparison of Citations Trends between the COVID-19 Pandemic and SARS-CoV, MERS-CoV, Ebola, Zika, Avian and Swine Influenza Epidemics

Artemis Chaleplioglou and Daphne Kyriaki-Manessi

*Department of Archival, Library & Information Studies, University of West Attica, Athens,*

*Greece*

Department of Archival, Library & Information Studies,

University of West Attica,

Egaleo Park Campus,

Ag. Spyridonos Str., Egaleo

Postal Code 12243, Athens, GREECE

E-mail: artemischaleplioglou@uniwa.gr

Tel. +30 210 5385100

Fax. 513-558-2269

**Abstract:**

**Objective:** The novel coronavirus COVID-19 outbreak rapidly evolved into pandemic. Global research efforts focus on this topic and with the collaboration of the scientific journals publication industry produced more than 16,000 related published articles in PubMed within five months from the onset of the outbreak. Herein, a comparison of the COVID-19 citations in PubMed and Web of Science was performed with SARS-CoV, MERS-CoV, Ebola, Zika, avian and swine influenza epidemics.

**Methods:** The citations were searched and collected using the disease terms and the date of publication restriction. The total number of PubMed citations and the HIV associated papers during the same chronological periods were examined in parallel. The journal category and country information of the publications were gathered from Web of Science. The collected data were statistically analyzed and compared.

**Results:** Significant correlations were found between COVID-19 and MERS (CC=0.988; p=0.003; q=0.006), Ebola (CC=0.987; p=0.003; q=0.011), and SARS (CC=0.964; p=0.015; q=0.028) epidemics five-month pick of novel citations in PubMed. However, COVID-19 publications were accumulated earlier and in larger numbers than any other 21$^{st}$ century major communicable disease outbreak.

**Conclusion:** The acceleration and the total number of COVID-19 publications represent an unprecedented landmark event in the medical library history. The immediate adoption of the fast-track peer-reviewing and publishing as well as the open access publication policies by the journal publishers are significant contributors to this bibliographic phenomenon.

**INTRODUCTION**

The emergence of novel communicable diseases represents a major global health threat despite the advances of modern biomedicine [1]. Human infectious disease outbreaks over time are rising globally with zoonoses to be the dominant kind [2]. Ecological catastrophies, climate change and global warming, population growth, migration, wars, lifestyle habits, poor public health conditions, microbial adaptation, globalization of travel and food supplies represent risk factors for the emergence, re-emergence and spread of infectious diseases [3]. The identification of relationships and the synthesis of knowledge networks between human diseases, diagnosis, molecular pathologies and therapies is one of the greatest challenges in medical library practice [4]. The citations trends could be used as markers to identify such correlations.

The novel coronavirus Severe Acute Respiratory Syndrome (SARS) Coronavirus-2 (SARS-CoV-2 or 2019-nCoV) that causes the coronavirus disease of 2019 (COVID-19) was initially identified in a cluster of fatal pneumonia cases in Wuhan, China [5]. With pathophysiological resemblance to SARS coronavirus, the first major epidemic of the new millennium [6, 7], and Middle East Respiratory Syndrome coronavirus (MERS-CoV) [8], COVID-19 outbreak was declared a pandemic on March 12, 2020 [9]. Many investigators shifted their research interests and were recruited in the fight against COVID-19. The emergency national responses to prevent transmission of the virus aimed to reduce interpersonal physical contact through quarantines and lockdowns. These actions affected many top research institutes and universities as well as publishers worldwide. However, many scientists continued their work, data analysis, collaboration and cooperation using modern technologies [10, 11]. The scientific journal publishing industry took immediate measures by revisiting the editorial priorities and scopes [12], fast tracking peer-review system or utilizing open peer-review platforms [13], accelerating accepted articles for online publishing, as well as by allowing open access to all the related to COVID-19 material. In parallel, international and national organizations provide multiple means and tools to facilitate COVID-19 research.

The worldwide collaboration of international organizations, governments, public bodies, research institutes, universities, databases resources and repositories, scientific journal publishers and scientists led to an outstanding accumulation of COVID-19 publications in a short period of time. Herein, the COVID-19 publication record and its features were compared to the publication output of previous major 21$^{st}$ century outbreaks of infectious diseases, in specific SARS-CoV, MERS-CoV, Ebola, Zika, avian and swine influenza epidemics. The aim of this work is to identify the publication trends and extract useful conclusions for medical library practice.

**METHODS**

**Search strategy**

Comprehensive PubMed and Web of Science (Core Collection) advanced searches were performed between May 11 and May 22, 2020. Searches were performed in PubMed for each of the terms "coronavirus COVID-19", "SARS", "Avian influenza or H5N1", "Swine influenza or H1N1", "MERS", "Ebola", or "Zika" included all fields in the query with date of publication restrictions. For each epidemic outbreak the search performed from the outbreak onset to two years after this date. Specifically, the chronological periods examined for each epidemic outbreak were from January 2020 to May 2020 for COVID-19 [5], January 2003 to December 2004 for SARS [14], January 2003 to December 2004 for avian influenza or H5N1 [15], December 2008 to November 2010 for swine influenza or H1N1 [16], September 2012 to August 2014 for MERS [17], December 2013 to November 2015 for Ebola haemorrhagic fever [18], and June 2015 to May 2017 for Zika microcephaly [19]. Total PubMed citations and "HIV or AIDS" related publications were also searched for January 2003 to May 2020. HIV and AIDS, representing a major life-threatening global virus disease epidemic from 1985 to 1995, literature was selected as control because of the regular trends of new publications overtime since the advent of antiretroviral drugs in 1996 that converted it into a "chronic disease" [20]. These searches were performed to identify regular PubMed new citations trends and changes per month and year. The same settings were applied in Web of Science searches with the terms searched as topics and by setting the timespan to custom year range to look for the respective chronological periods.

**Data collection, Journals, Categories and Countries**

The citations retrieved from PubMed or Web of Science were collected either as comma-separated values (.csv) or text (.txt) files. All the citations collected were included in the analysis without exceptions or exclusions. All data were transferred in spreadsheets (.xls or

.xlsx) for further analysis organized by the disease and the month of article publication. Total PubMed citations and "HIV or AIDS" related publications were also organized by the month of publication. For the citations information collected from Web of Science data were organized by disease and journal category or country. For the journal category 0.5% of COVID-19 and 0.2% of swine influenza citations were not informative. For the country information data collected from Web of Science, 3% of COVID-19, 19% of SARS, 13% of avian influenza, 11% of swine influenza, 6% of MERS, 30% of Ebola, and 16% of Zika citations were not informative. The Clarivate Journal Citation Reports (JCR) Science Editions and ScimagoJR reports for the years 2003, 2009, 2012, 2014, 2016 and 2019 were used to collect the number of indexed scientific journals.

**Statistical analysis**

Descriptive statistics, multiple regression, linear regression, calculation of slope, polynomial nonlinear regression, Kolmogorov-Smirnov normality tests, Pearson correlation coefficients and Student t-test were performed with Microsoft (MS) Excel and SPSS SigmaStat. Data are presented as mean ± standard deviation. Normality tests revealed that all data were derived from a population with a normal distribution. A p-value of <0.05 was considered statistically significant. However, to reduce multiple correlation comparisons false discovery rate (FDR) the Dunn-Sidak correction was applied and the q-values were determined. A q-value of <0.05 was considered statistically significant. Power calculations were performed, on the basis of the weakest associations. The sample size testing in this study would allow the identification of correlation of 0.95 with a power of 99.8% at the 0.05 significance level. Two-, three- or four-set Venn diagrams showing the logical relations between datasets were calculated and designed.

**RESULTS**

**General Trends of Scientific Journals and Publications**

The PubMed citations trends for the chronological periods of the epidemics studied were explored (Figure 1). The average monthly new publications found to be increased from 54,332.9 citations/month in 2003 to 148,712.4 citations/month in 2019 (Figure 1A). This is an indicator of an increase scientific output as well as of an increase in scientific journals during the past 17 years. Indeed the scientific journals indexed in JCR and ScimagoJR found to be increased by 110% and 80% respectively (Figure 1B and 1C). Differences in the slope of new publications in PubMed per month were observed in the years of interest (Figure 1D). This finding suggests the need of bibliographic controls when compare the citation trends of different topics in different chronological periods. The controls normalize the data according to the general trends that affect the corpus of scientific journal publications.

The behavior of the PubMed citations with time through the years 2002-2020 (Figure 2A), and with the month of publication release were examined (Figure 2B). Linear regression model reveal an $R^2$ of 0.992 goodness-of-fit of PubMed citations for the time period analyzed. Significant fluctuations and trends were identified when the PubMed citations were analyzed by the month of publication (Figure 2B). The new publications in January are systematically 60% more than the yearly average of new publications per month because of the new issues of annual journals editions, while the new publications in February and August are 10.8% and 11.5% less, respectively. The deviation of new publications in month January through the years tested is 24%, while the average standard deviation for the rest months of the year is 32%.

The monthly output of PubMed citations is numerically substantially greater than the monthly citations related to the outbreak epidemics examined, on average 10,000 times more. Therefore, HIV or AIDS related publications in the same chronological periods to the epidemic outbreaks were postulated as normalizing controls. AIDS is the first postmodern pandemic [21] with 16 titles of subject specific associated journals in JCR index and 29 titles in ScimagoJR 2018 editions. The monthly output of HIV or AIDS related citations in PubMed is numerically closer to the monthly citations related to the outbreak epidemics examined, on average 53 times higher. The HIV or AIDS related PubMed citations were explored with time for the years 2002-2020 (Figure 2C), and with the month of publication release (Figure 2D). A goodness-of-fit $R^2$ of 0.878 with linear regression was found while significant similarities with the total new PubMed citations trends by the year of publication or month were observed. In specific, the new HIV or AIDS publication for Januray is 67% more than the annual average, with a deviation of 30% through the years, while February and August citations are less by 20% and 11% compared to the annual average, respectively, and the average monthly citations deviation, with the exception of January, is 21.6%. These similarities were confirmed with the Pearson correlation coefficient of 0.955 (Student t-test, 2-tail, p=0.017) between HIV or AIDS citations to the total new PubMed citations. Thus, the HIV or AIDS citations of the same time of interest could be used as the normalizing control of bibliographic data.

**Trends of COVID-19 Citations**

The COVID-19 related PubMed citations trends for the last five months were analyzed (Figure 3). The new COVID-19 citations together with the HIV or AIDS related new publications per week are depicted in Figure 3A, for the chronological period of December 29, 2019 to May 24, 2020. The absolute slope of new COVID-19 citations in PubMed per week, calculated by linear regression, was found to be 130 and the relative slope 17%. The new COVID-19 publications exceeded one-hundred threshold during the week February 3 to February 9, 2020, three weeks post the outbreak, and one-thousand during the week April 6 to April 12, 2020, fifteen weeks post the outbreak. They surpass HIV or AIDS new citations for the first time during the week March 16 to March 22, 2020, while one week later the difference was 2-fold, and four weeks later 3-fold. The difference between COVID-19 and HIV or AIDS new citations is up to this date continuously growing to 8-fold (May 24, 2020). When compared to the total number of new PubMed citations per week for the same chronological period, COVID-19 new citations reached 1% of the total during the week March 2 to March 8, 2020, 10 weeks post the outbreak, exceeded 5% by April 13 to April 19, 2020, sixteen weeks post the outbreak, and 10% of the total by May 18 to May 24, 2020, 21 weeks post the outbreak. This finding strongly suggests a dynamic shift in scientific research efforts and journal scopes of publication.

These trends are also evident in the rate of accumulation of COVID-19 citations in PubMed per week (Figure 3B). The slope of the linear regression curve is almost 800 (relative 19.8%; $R^2$=0.795). The polynomial nonlinear regression trendline ($R^2$=0.989) reveals an inflection point in the fifteenth week post the outbreak, the week between April 6 and April 12, 2020, when the total citations were approximately 4,000, a quarter of the number of citation to May 24, 2020 . From this point forward the accumulation of COVID-19 citations is increasing sharply with time.

To explore for any consequences of the COVID-19 pandemic, especially the emergency national responses against it, in the scientific journals publishing, an analysis of the new citations per month for the particular months January, February, March and April was performed for the years 2015 to 2020 (Figure 3C). It is evident that the general trends of publication changes with time for each month were not affected by the COVID-19 pandemic at least in its early phase. The relative slope of citations for January for the years 2015-2019 was found to be -3.9% compared to -3.8% when January 2020 is included. The relative slopes for February, March and April 2015-2019 found to be 4.6%, 2.9% and 2.8%, respectively, compared to 4.8%, 3.1% and 3.7% when the corresponding months of 2020 are included. So far, it seems that globalization and remote access to labor effectively retain the operational capabilities of the scientific publication industry. However, it remains to be elucidated in the future whether the restriction of scientific research laboratory work for several months in many top institutes and universities will ultimately affect the novel citation incorporation rate in PubMed.

**Comparison of COVID-19 Literature to 21$^{st}$ Century Communicable Disease Outbreaks**

Comprehensive comparisons of 21st century communicable disease epidemics trends of related publications from the initial outbreak to two years afterwards were performed (Figure 4). The total number of relative publications at the end point of two years post the outbreak found to be for SARS 2543, for Avian influenza or H5N1 498, for Swine influenza or H1N1 10411, for MERS 502, for Ebola 3700, for Zika 3645, and for COVID-19 till May 24, 2020 16213 citations. The slopes for linear regression curve of citations accumulation in PubMed per month were found to be for SARS 114 (relative 8,8%), for Avian influenza or H5N1 17.5 (6.7%) 498, for Swine influenza or H1N1 465 (9.5%), for MERS 22 (11.7%), for Ebola 181 (12.4%), for Zika 176 (13.3%), and for COVID-19 till May 24, 2020 4080 (70%). The comparison of the accumulation curves of related articles in PubMed for 24 months after each outbreak epidemic clearly shows the markedly increase of COVID-19 publications in short time after its outbreak (Figure 4A).

When the normalized citation data of related epidemic publications by the HIV or AIDS corresponding chronological publications were examined, COVID-19 literature demonstrates a unique dynamical increase versus the rest of the 21st century epidemics (Figure 4B). The new publications of COVID-19 surpass the regular increase of HIV or AIDS relative literature by 3.5- to 14-folds from March to May, starting from 8% in January to 87% in February, 346% in March, 940% in April, to 1305% in May 2020. For the rest epidemics the maximum ratio of relative citations to HIV or AIDS found to be for SARS 63.3% within 5 months (median of 28% achieved in 5 months), for Avian influenza or H5N1 18.8% within a month (median 5% in 1 month), for Swine influenza or H1N1 120% within 13 months (median 87.7% in 9 months), for MERS 6.7% within 21 months (median 3% in 12 months), for Ebola 47% within 15 months (median 25.8% in 11 months), and for Zika 35.6% within 15 months post the outbreak (median 28.6% in 10 months). Significant deviations from a linear curve were observed for all normalized by HIV or AIDS ratios of related to epidemic citations with time with $R^2$ of 0.479 for Swine influenza or H1N1, 0.004 for SARS, 0.507 for Ebola, 0.698 for Zika, 0.009 for Avian influenza or H5N1, and 0.670 for MERS epidemics. Only the COVID-19 normalized by HIV or AIDS ratios of citations exhibited a goodness-of-fit $R^2$ of 0.934 with linear regression, an indicator of a shift for a significant part of the scientific research community towards COVID-19.

Since the history of COVID-19 pandemic is restricted to the past 5 months, the 5-month slopes of new relative to the epidemics publications were analyzed on the basis of 5-month overlaid periods post the outbreak (Figure 4C). Irregular intensity of the scientific research efforts for all epidemics except COVID-19 is also apparent in this diagram. The slope for the first 5-month of COVID-19 pandemic is the highest. The best 5-month slopes of relative publication increase have been achieved in different chronological periods from the onset of each epidemic. In specific, for SARS on month 6 post the outbreak, for Avian influenza or H5N1 on month 15, for Swine influenza of H1N1 on month 14, for MERS on month 13, for Ebola haemorrhagic fever on month 12, and for Zika microcephaly on month 11. Despite the irregularities Kolmogorov-Smirnov normality test indicates that the data matches to the pattern expected from a population with a normal distribution. The best 5-month slopes of relative publications increase for each epidemic dataset were used in Pearson correlation analysis (Figure 4D). Statistical significant correlations (Student t-test p<0.05) were observed between COVID-19 and MERS (Correlation coefficient 0.988), COVID-19 and Ebola (CC 0.987), MERS and Ebola (CC 0.985), MERS and Zika (CC 0.972), COVID-19 and SARS (CC 0.964), Ebola and Zika (CC 0.961), SARS and MERS (CC 0.958), SARS and Ebola (CC 0.945) and COVID-19 and Zika (CC 0.937). Due to multiple correlations testing the false discovery rate (Dunn-Sidak) applied and the COVID-19 and Zika correlation was rejected (q>0.05). The correlation data suggest similar trends of publication record increase during these 5-month periods.

**Comparison of Scientific Journal Categories and Countries of Publications**

To understand the relationships between COVID-19 and the rest 21$^{st}$ century communicable outbreak epidemics the journal categories according to Web of Science that published the relative literature were explored. For COVID-19 literature 174 different journal categories were retrieved compared to 211 for SARS, 64 for Avian influenza or H5N1, 135 for Swine influenza or H1N1, 41 for MERS, 123 for Ebola, and 160 for Zika epidemics. Medicine general internal is the predominant category of the publication records of COVID-19, Ebola and SARS, infectious diseases category of Swine influenza or H1N1, MERS and Zika, and veterinary sciences of Avian influenza or H5N1. The journal category data were compared in two-way Venn diagrams to identify the intersections of the epidemics literature lists. The COVID-19 journal category list found to contain all, or nearly all, the categories of MERS (100%), Ebola (87.8%), the Avian influenza or H5N1 (90.6%) and the Swine influenza or H1N1 (92.6%), as well as to exhibit similarities with Zika (86.9%) and SARS (72%). The differences between COVID-19 and SARS journal category lists was an unexpected finding because of the taxonomic similarities of the two viruses [5]. The particular journal categories that exist only in COVID-19 publication record compared to SARS were related to the direct or indirect consequences of the lockdowns, administrative, environmental, social, financial, psychological and emotional, including public administration, civil engineering, acoustics, industrial relations and labor, philosophy, religion, psychology and psychoanalysis, and substance abuse. The journal categories that exist only in SARS publication record and not in COVID-19 are related to the geographic distribution of SARS epidemic and the research for the animal carrier of the disease. The most prevalence journal categories for COVID-19, MERS, Ebola, and SARS, the diseases that exhibited good correlated 5-month slope of new publications are presented in Figure 5. A four-way Venn diagram demonstrated the close relationships between COVID-19, MERS and Ebola, as well as the differences with SARS.

The distribution of the countries of origin of the publications of the epidemics was also studied. COVID-19 publications originated from 114 different countries worldwide, while SARS from 66, Avian influenza or H5N1 from 36, Swine influenza or H1N1 from 85, MERS from 27, Ebola from 79 and Zika from 131. Comparisons of the countries of origin were performed in Venn diagrams. All countries that published MERS (100%) and Avian nfluenza or H5N1 papers (100%) and nearly all of the countries that published SARS (88%), Swine influenza or H1N1 (86%), Ebola (84%) and Zika (72%) have also published COVID-19 papers. The countries that have been issued Zika papers and not COVID-19 include mostly Caribbean and African countries. Most of COVID-19 publications are originated from Peoples Republic of China, where the epidemic started, followed by USA, Italy, England, India and Iran. Most of SARS epidemic related publications are also came from Peoples Republic of China, followed by USA, Canada, Taiwan, Singapore, and Germany. For all the rest 21$^{st}$ century epidemics most papers came from USA, followed for Ebola by England, Canada and France, for MERS by the People's Republic of China, Saudi Arabia, Germany and England, for Avian influenza or H5N1 by the People's Republic of China, England, Italy and the Netherlands, for Swine influenza or H1N1 by the People's Republic of China, England, Germany, Canada, Japan and Australia, and for Zika by Brazil, France and England. The distributions of the countries of origin of the related publications for COVID-19, MERS, Ebola and SARS are depicted as pie-charts in Figure 6. A four-way Venn diagram exhibits the similarities and differences between these four epidemics by the countries of origin of publications. COVID-19 exhibits more common countries with Ebola epidemic followed by SARS and MERS epidemics. COVID-19 publications records include documents from 35 countries that have not contributed as countries of origin for SARS, Ebola or MERS. This list includes Asian countries, Afghanistan, Nepal, Bahrain, Kazakhstan, Georgia, Kuwait, Indonesia, Sri Lanka, Lebanon and Iraq, as well as European countries, like Belarus, Bulgaria, Cyprus, Romania, Slovenia, and Malta, African countries, like Ethiopia, Morocco, Tunisia, Mozambique and Sudan, and Caribbean and Latin America countries, such as Costa Rica, Honduras, Trinidad, Grenada, Ecuador and Paraguay.

**DISCUSSION**

Comparison of epidemics is an important scientific tool for the comprehension of novel emerging or remerging communicable diseases. Data sources, quantification of trends and elimination of biases are key factors in this process [1]. Herein, the recent scientific literature on COVID-19 pandemic until May 2020 was bibliographically quantitative compared to the relative to the rest 21$^{st}$ century major epidemics, in specific SARS, Avian influenza, Swine influenza, MERS, Ebola and Zika, that was published within two years after the outbreaks. The analysis was based on PubMed citations and publication dates controlled by the total PubMed citations and the AIDS related literature at the corresponding chronological intervals. The fluctuations of publication issued over time was used as an indicator of the scientific research community and the journal publication industry response against the epidemics. Emphasis was given in the time from the epidemic outbreak to the pick increase of literature volume. The COVID-19 literature was rapidly expanded in numbers but correlations with MERS, Ebola and SARS were identified. The journal categories and the countries of origin of the citations was extracted from Web of Science, and was used to assess the specific scopes of scientific research on the particular diseases during the epidemics as well as the geographical distribution of research. The differences of COVID-19 related journals research scopes to the rest epidemics were attributed to the worldwide application of lockdowns as emergency measures. The geographical differences were associated to the patters of disease spread of the epidemics.

Analyses of the literature of 21$^{st}$ century communicable disease epidemics have been previously performed for review, historical, bibliographic or bibliometric perspective, for SARS [22], for MERS [23], for SARS and MERS [24], for Avian influenza [25], for Swine influenza [26], for Ebola [27], for Zika [28, 29], for Zika and Ebola [30]. To the best to our knowledge this is the first report of a comprehensive correlation analysis of the bibliographic trends of all SARS, Avian influenza, Swine influenza, MERS, Ebola and Zika epidemics. Moreover, in this report the whole scientific literature corpus related to these epidemics for the chronological intervals studied retrieved from PubMed and Web of Science and analyzed. Significant correlation of the increase of publications with time post the outbreak were found between MERS and Ebola, MERS and Zika, Ebola and Zika, SARS and MERS, as well as SARS and Ebola. These similarities reflect a similar pattern of scientific research behavior but not at the same extend regarding the total number of publications or the pick value of new publications. The geographical distribution spread of the disease, the difference between pandemic and epidemic, and the fatality risk severely affect the total number of citations as well as the pick number of new articles per month. SARS, Avian influenza, MERS, Ebola and Zika epidemics appear to be geographically localized and contained when compared to Swine influenza and COVID-19 pandemics [31-33]. We found that COVID-19 publication record already at 16,000 citations in five months and Swine influenza record 10,000 citations two years post the outbreak of the pandemic, are by far more extended than Ebola with 3,700, Zika with 3,600, or SARS with 2,500, all documents published within two years post the initial outbreak. AIDS citations published during the same chronological periods to the epidemics and pandemic, were used as controls since from 1996 it is consider a "chronic disease" [20], AIDS citations correlated well with the total PubMed citations, they followed the same regular distributions and variations of new citations per specific month of the year to the total PubMed record, and the size of their average number is of the same magnitude with the literature of the epidemics and pandemics studied. When the normalized by AIDS publications epidemics literature were analyzed irregular changes of their ratios were observed. Comparison of the normalized data with COVID-19 clearly demonstrated the uniqueness and the worldwide impact of this disease, similar to the initial pandemic impact of AIDS [21].

The COVID-19 literature trends suggests that worldwide scientific collaboration, flexibility in scientific research purposes and aims, and fast-track peer-reviewing and open access publication policies by the publishers may impressively shift the community towards a new critical priority, an outbreak emergence call, to produce results and propose solutions. Future studies will assess the bibliometric impact of COVID-19 published citations with an appropriate perspective.

**Figure Legends**

**Figure 1.** General Trends of Scientific Journals and Citations. (A) The average new PubMed publications for the chronological periods of 21st century epidemic outbreaks are presented. A substantial increase of citations with time is observed. (B) This increase is partly due to the increase of scientific journals. The JCR scientific indexed journals with time are is presented. (C) The ScimagoJR indexed journals with time are presented. (D) The slopes of the increase of new publications in PubMed per month are presented. The differences in the slopes are indicators of changes that affected globally the medical scientific journals publication industry.

**Figure 2.** Periodical Fluctuations of Publications per Month. (A) Diagrammatic presentation of the new publications in PubMed per year between 2002 and 2019. The resulting linear regression trendline is depicted with a dashed line, together with its equation and R2. (B) Diagrammatic presentation of the average new PubMed publications per month, as bars with the standard deviation for each month for all years between 2002 and 2020. (C) Diagrammatic presentation of the new HIV or AIDS related publications in PubMed per year between 2002 and 2019. The resulting linear regression trendline is depicted with a dashed line, together with its equation and R2. (D) Diagrammatic presentation of the average new HIV or AIDS related publications in PubMed per month, as bars with the standard deviation for each month for all years between 2002 and 2020. The fluctuations of new HIV or AIDS related publications are well correlated with the total new PubMed publications.

**Figure 3.** COVID-19 citations trends. (A) The new COVID-19 and HIV or AIDS related publications in PubMed per week for the chronological period of December 29, 2019 to May 24, 2020 are presented. New COVID-19 publications surpass HIV or AIDS publication for the first time during the week March 16 to March 22, 2020. This difference is up to this date continuously increasing to almost 8-times. (B) The accumulation of COVID-19 publications in PubMed per week is documented diagrammatically. The resulting linear and polynomial regression trendlines are depicted with dashed lines, purple and red respectively, together with the equations and R2. (C) January, February, March and April 2020 total PubMed new publications are depicted in comparison with the same months for years 2015-2019. The publication trends of 2020 follow similar patterns of decrease for January and increase for months February, March and April. COVID-19 early phase pandemic didn't affect the scientific journal publication industry.

**Figure 4.** Comprehensive comparisons of 21st century communicable disease epidemics trends of related publications from the initial outbreak to two years afterwards. (A) Diagrammatic presentation of the total number of related articles in PubMed after the outbreak with time. COVID-19 exhibits a sharp increase of publications that surpass any other 21st century epidemic. (B) Diagrammatic presentation of the number of new related publications per month post the outbreak normalized by HIV or AIDS citations of the matched chronological period. New publications on COVID-19 rapidly surpass the regular HIV or AIDS relative literature by 4- to 14-times. (C) Diagrammatic presentation of 5-month slope of new relative to the epidemics publications in 5-month periods post the outbreak. The exact chronological periods that the pick slope values achieved are depicted with matching color to the line of its epidemic. The level of the COVID-19 5-month slope is marked in dark red dashed line and the chronological period occurred is depicted in the same color. The first 5-month slope of COVID-19 is higher than any other pick value of the rest 21st century communicable epidemics. (D) The results of Pearson correlations

coefficients are presented with the Studend t-test, 2-tailed, p-values. When a p-value is <0.05, the false discovery rate q-value is also presented, calculated according to Dunn-Sidak. COVID-19 5-month slope found to be correlated well with MERS, Ebola, and SARS epidemics.

**Figure 5.** Comparison of the journal categories of COVID-19, MERS, Ebola and SARS publications according to Web of Science categorization. COVID-19 exhibits more common journal categories with SARS than by MERS and Ebola.

**Figure 6.** Comparison by the country of origin of COVID-19, MERS, Ebola and SARS publications according to Web of Science. COVID-19 exhibits more common countries with Ebola epidemic followed by SARS and MERS epidemics.

**FIGURES**

**Figure 1**

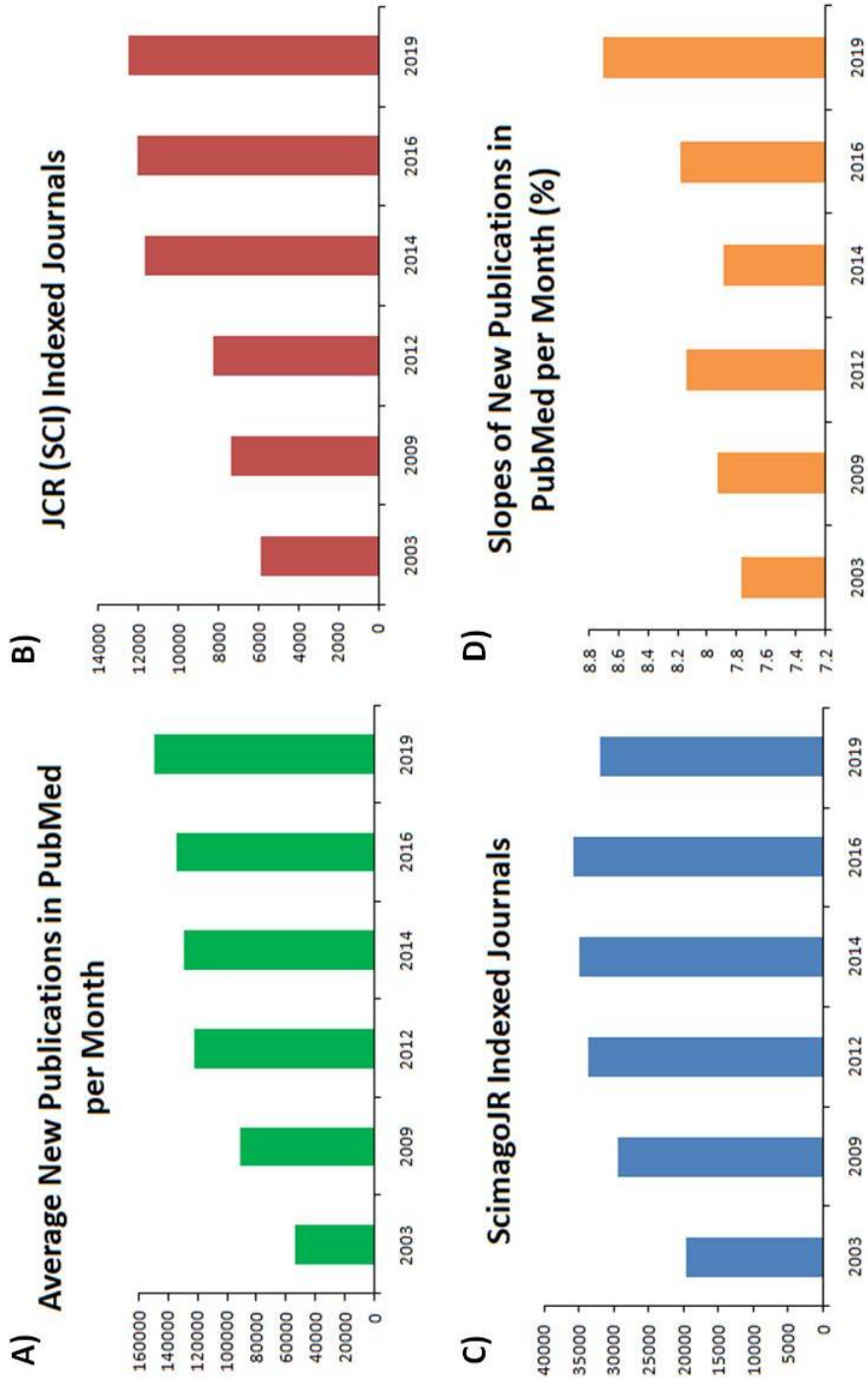

**Figure 2**

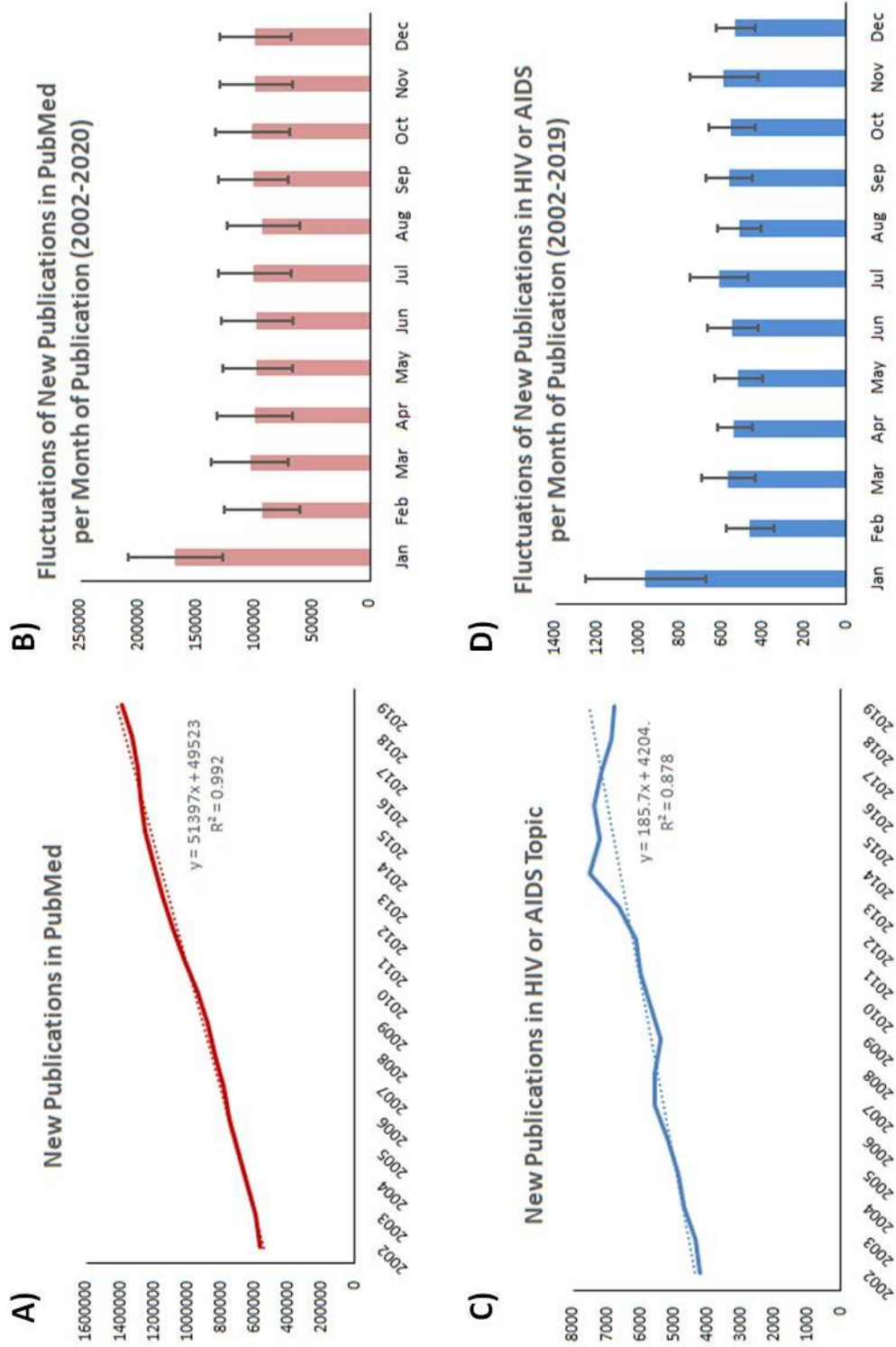

**Figure 3**

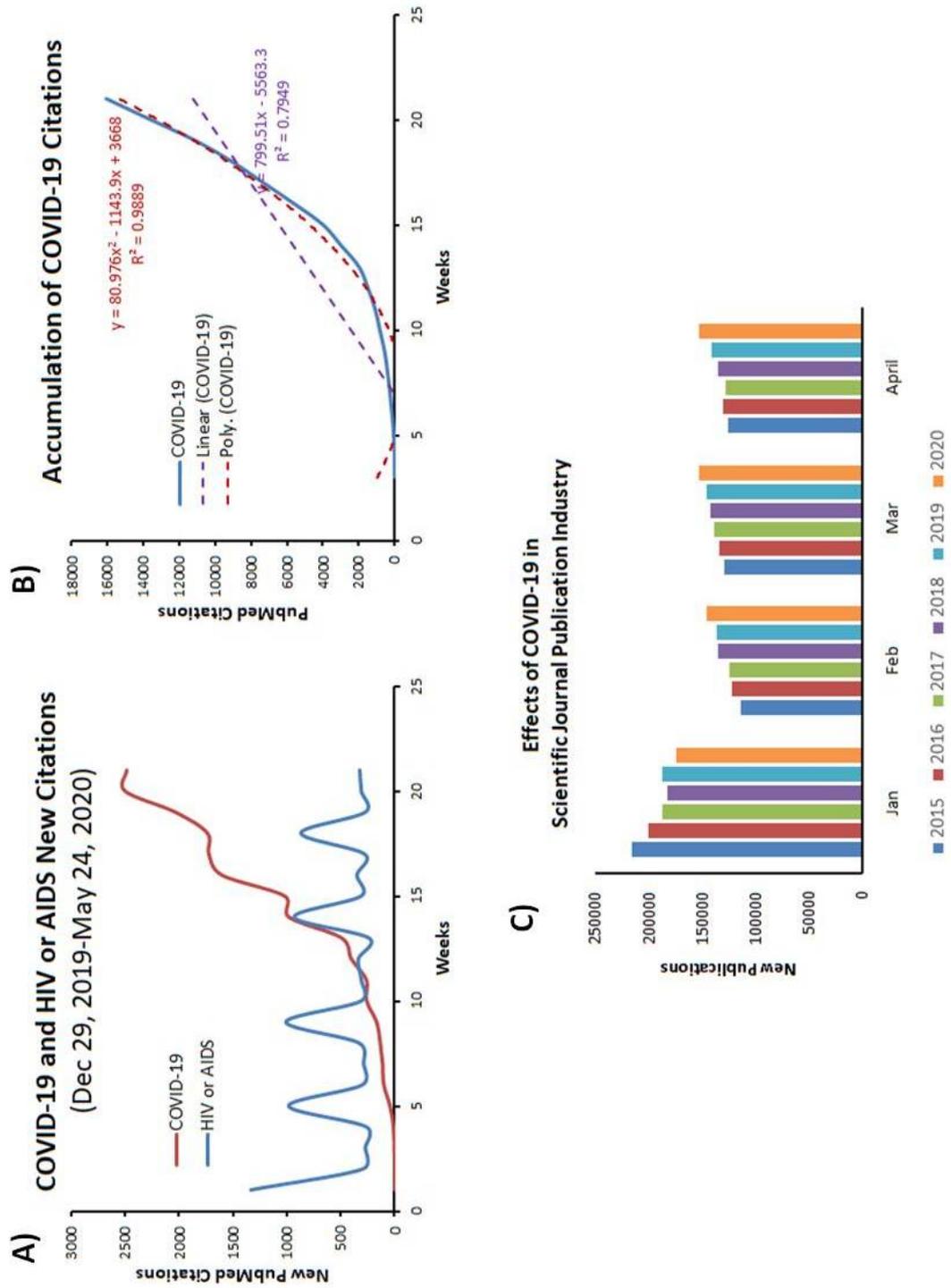

**Figure 4**

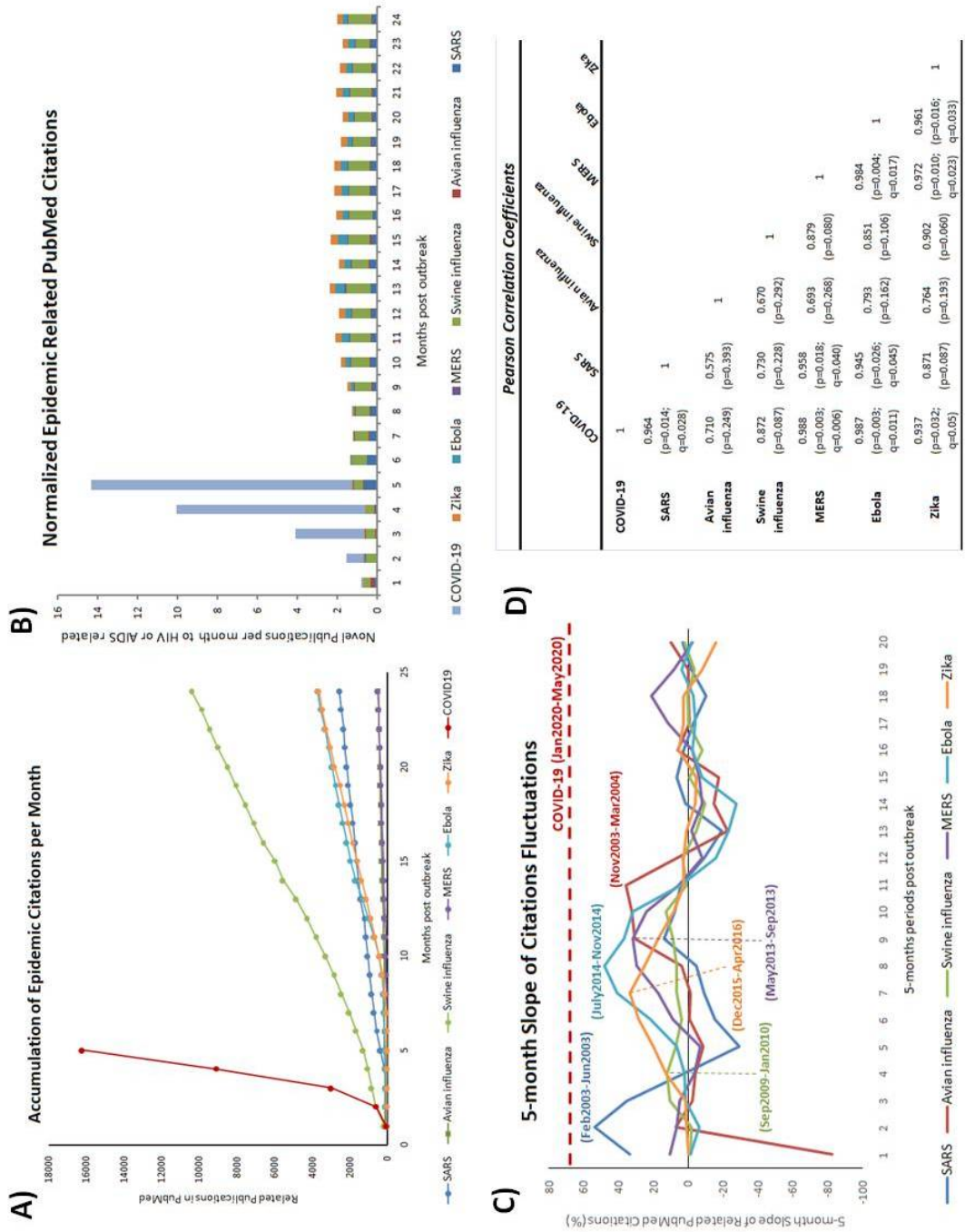

**Figure 5**

**Figure 6**

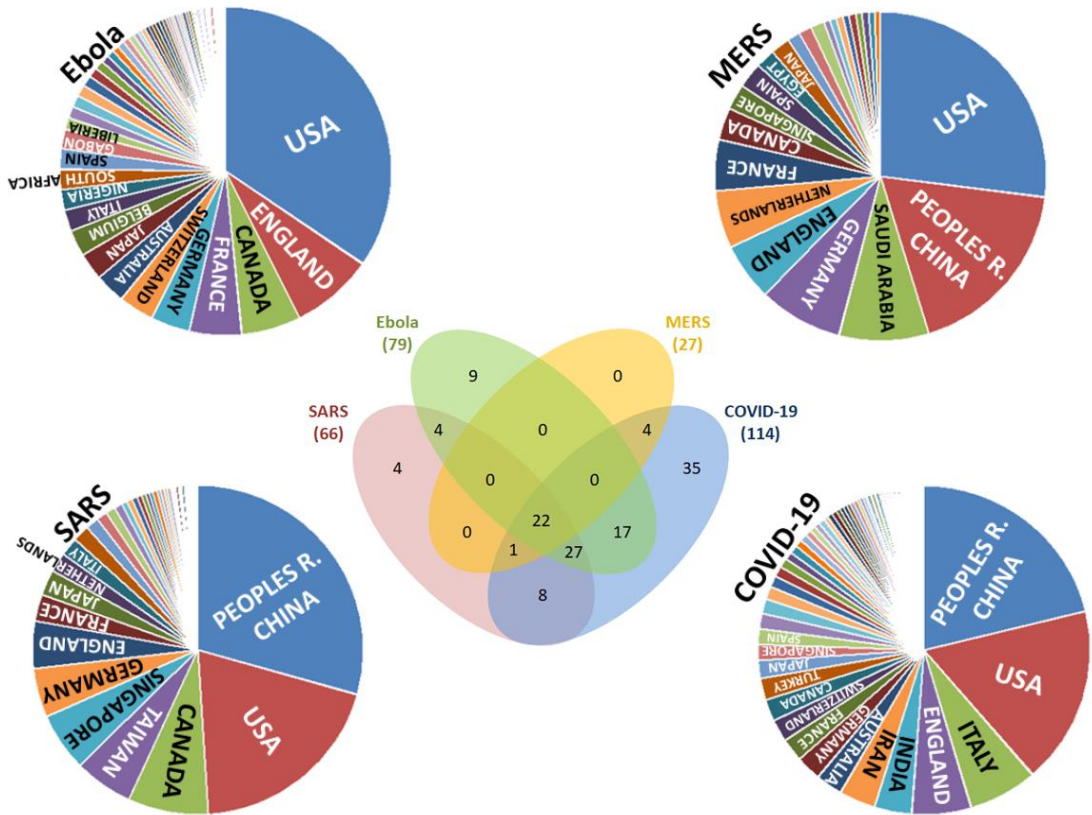